\documentclass[aps,pre,preprint,groupedaddress,showpacs]{revtex4}
\usepackage{graphicx}
\usepackage{amsmath}

\begin{document}

\title{Power laws and self-similar behavior in negative ionization fronts}
\author{Manuel Array\'as$^{1}$, Marco A. Fontelos$^{2}$ and Jos\'e L. Trueba$^{1}$}
\affiliation{$^{1}$Departamento de Matem\'aticas y F\'{\i}sica
Aplicadas y Ciencias de la Naturaleza, Universidad Rey Juan
Carlos, Tulip\'{a}n s/n, 28933 M\'{o}stoles, Madrid, Spain}
\affiliation{$^{2}$Departamento de Matem\'aticas, Universidad
Aut\'onoma de Madrid, 28049 Cantoblanco, Madrid, Spain}

\begin{abstract}
We study anode-directed ionization fronts in curved geometries.
When the magnetic effects can be neglected, an electric shielding
factor determines the behavior of the electric field and the
charged particle densities. From a minimal streamer model, a
Burgers type equation which governs the dynamics of the electric
shielding factor is obtained. A Lagrangian formulation is then
derived to analyze the ionization fronts. Power laws for the
velocity and the amplitude of streamer fronts are observed
numerically and calculated analytically by using the shielding
factor formulation. The phenomenon of geometrical diffusion is
explained and clarified, and a universal self-similar asymptotic
behavior is derived.
\end{abstract}

\date{\today}


\maketitle

\section{Introduction}
In a perfect dielectric medium, charged particles form
electrically neutral atoms and molecules due to powerful electric
forces. Since there are not free charges in this medium, electric
current does not flow inside it. However, if a very strong
electric field is applied to a medium of low conductivity in such
a way that some electrons or ions are created, then mobile charges
can generate an avalanche of more charges by impact ionization, so
that a low temperature plasma is created, and an electric
discharge develops \cite{Rai}. This process is called electric
breakdown and it is a threshold process: there are not changes in
the state of the medium while the electric field across a
discharge gap is gradually increased, but at a certain value of
the field a current is created and observed.

A streamer is a ionization wave propagating inside a non-ionized
medium, that leaves a non-equilibrium plasma behind it. They
appear in nature and in technology \cite{Rai,Eddie}. A streamer
discharge can be modelled using a fluid approximation based on
kinetic theory \cite{guo}. Defining the electron density $N_{e}$
as the integral of the electron distribution function over all
possible velocities, we get
\begin{equation}
\frac{\partial N_{e}}{\partial \tau} + \nabla_{\bf R} \cdot {\bf
J}_{e} = S_{e}, \label{balance-electrons}
\end{equation}
where $\tau$ is the physical time, $\nabla_{\bf R}$ is the
gradient in configuration space, ${\bf U}_{e} ({\bf R}, \tau)$ is
the average (fluid) velocity of electrons, $S_{e}$ is the source
term, i.e. the net creation rate of electrons per unit volume as a
result of collisions, and ${\bf J}_{e} ({\bf R}, \tau) = N_{e}
({\bf R}, \tau)\, {\bf U}_{e} ({\bf R}, \tau)$ is the electron
current density. Similar expressions can be obtained for positive
$N_{p}$ and negative $N_{n}$ ion densities.

A usual procedure is to approximate the electron current ${\bf
J}_e$ as the sum of a drift (electric force) and a diffusion term
\begin{equation}
{\bf J}_e=-\mu_e {\boldsymbol{\cal E}} N_e - D_e \nabla_{\bf R} N_e ,
\label{el-current}
\end{equation}
where ${\boldsymbol{\cal E}}$ is the total electric field (the sum
of the external electric field applied to initiate the propagation
of a ionization wave and the electric field created by the local
point charges) and $\mu_e$ and $D_e$ are the mobility and
diffusion coefficients of the electrons. Note that, as the initial
charge density is low and there is no applied magnetic field, the
magnetic effects in equation (\ref{el-current}) are neglected.
This could not be done in cases where the medium is almost
completely ionized or cases in which an external magnetic field is
applied, leading to different treatments \cite{Chandra}.

Some physical processes can be considered giving rise to the
source terms. The most important of them are impact ionization (an
accelerated electron collides with a neutral molecule and ionizes
it), attachment (an electron may become attached when collides
with a neutral gas atom or molecule, forming a negative ion),
recombination (a free electron with a positive ion or a negative
ion with a positive ion) and photoionization (the photons created
by recombination or scattering processes can interact with a
neutral atom or molecule, producing a free electron and a positive
ion) \cite{liu}.

It is also necessary to impose equations for the evolution of the
electric field ${\boldsymbol{\cal E}}$. It is usual to consider
that this evolution is given by Poisson's law,
\begin{equation}
\nabla_{\bf R}\cdot{\boldsymbol{\cal E}} = \frac{e}{\varepsilon_{0}} \,
\left( N_{p} - N_{n} -N_{e} \right) , \label{poisson}
\end{equation}
where $e$ is the absolute value of the electron charge,
$\varepsilon_{0}$ is the permittivity of the gas, and we are
assuming that the absolute value of the charge of positive and
negative ions is $e$.

Some simplifications can be made when the streamer development out
of a macroscopic initial ionization seed is considered in a
non-attaching gas like argon or nitrogen. For such gases,
attachment, recombination and photoionization processes are
usually neglected. A minimal model turns out and has been used to
study the basics of streamer dynamics
\cite{DW,Vit,Ute,ME,ME1,Andrea}. In those cases the evolution of
electron and positive ion densities in early stages of the
discharge can be written as
\begin{eqnarray}
\frac{\partial N_{e}}{\partial \tau} &=& \nabla_{\bf R} \cdot \left( \mu_e {\boldsymbol{\cal E}} N_e + D_e
\nabla_{\bf R} N_e \right) + \nu_{i} N_e,
\label{model01} \\
\frac{\partial N_{p}}{\partial \tau} &=& \nu_{i} N_e. \label{model02}
\end{eqnarray}
On time scales of interest the ion current is more than two orders
of magnitude smaller than the electron one so it is neglected in
(\ref{model02}). In these equations $\nu_{i} N_e$ is a term
accounting for impact ionization, in which the ionization
coefficient $\nu_{i}$ is given by the phenomenological Townsend's
approximation,
\begin{equation}
\nu_{i} = \mu_e |{\boldsymbol{\cal E}}| \alpha_0 e^{- {\cal E}_0 /|{\boldsymbol{\cal E}}|}, \label{townsend}
\end{equation}
where $\mu_e$ is the electron mobility, $\alpha_{0}$ is the
inverse of ionization length, and ${\cal E}_{0}$ is the
characteristic impact ionization electric field.

It is convenient to reduce the equations to dimensionless form.
The natural units are given by the ionization length
$R_0=\alpha_0^{-1}$, the characteristic impact ionization field
${\cal E}_0$, and the electron mobility $\mu_e$, which lead to the
velocity scale $U_0=\mu_e {\cal E}_0$, and the time scale
$\tau_0=R_0/U_0$. The values for these quantities for nitrogen at
normal conditions are $\alpha_0^{-1} \approx 2.3\,\mu\mathrm m$,
${\cal E}_0 \approx 200$ \,kV/m, and $\mu_e \approx 380\,\mathrm
{cm^2/(Vs)}$. We introduce the dimensionless variables ${\bf
r}={\bf R}/R_0$, $t=\tau/\tau_0$, the dimensionless field ${\bf
E}={\boldsymbol{\cal E}}/{\cal E}_0$, the dimensionless electron
and positive ion particle densities $n_e=N_e/N_0$ and
$n_p=N_p/N_0$ with $N_0=\varepsilon_0 {\cal E}_0/(e R_0)$, and the
dimensionless diffusion constant $D=D_e/(R_0 U_0)$.

In terms of the dimensionless variables, the minimal model
equations become
\begin{eqnarray}
\label{1}
\frac{\partial n_e}{\partial t} &=& \nabla\cdot {\bf j} + n_e f(|{\bf E}|),\\
\label{2}
\frac{\partial n_p}{\partial t} &=& n_e f(|{\bf E}|), \\
\label{3}
n_p - n_e &=& \nabla\cdot{\bf E}, \\
\label{4}
{\bf j} &=& n_e {\bf E} + D\; \nabla n_e , \\
\label{ft} f(|{\bf E}|) &=& |{\bf E}| e^{-1/|{\bf E}|},
\end{eqnarray}
where $\nabla = \nabla_{\bf r}$, and ${\bf j}$ is the
dimensionless electron current density.

Some properties of planar fronts have been obtained analytically
\cite{Ute,pulled1,AJP} for the minimal model. A spontaneous
branching of the streamers from numerical simulations has been
observed \cite{ME}, as it occurs in experimental situations
\cite{Pasko}. In order to understand this branching, the
dispersion relation for transversal Fourier-modes of planar
negative shock fronts (without diffusion) has been derived
\cite{ME1}. For perturbations of small wave number $k$, the planar
shock front becomes unstable with a linear growth rate
$|E_\infty|k$. It has been also shown that all the modes with
large enough wave number $k$ (small wave length perturbations)
grow at the same rate (it does not depend on $k$ when $k$ is
large). However, it could be expected from the physics of the
problem that a particular mode would be selected. To address this
problem, a possibility is to consider the effect of diffusion. It
is also interesting to investigate the effect of electric
screening since, in the case of curved geometries, this screening
might be sufficient to select one particular mode.

In this paper, we study the properties and structure of
anode-directed ionization fronts with zero diffusion coefficient
for curved geometries. We start discussing the consequences of
neglecting the magnetic field effects in the physics of the
streamer evolution. As a consequence, an electric shielding factor
can be introduced which determines the behavior of the electric
field and the particle densities. From the minimal streamer model,
a Burgers type equation which govern the dynamics of the electric
shielding factor is deduced. This allows us to consider a
Lagrangian formulation of the problem simplifying the analytical
and numerical study of the fronts. We apply this new formulation
to planar as well as curved geometries (typical in experimental
set-ups). Power laws for the velocity and the amplitude of
streamer fronts are observed numerically. Theses laws are also
calculated analytically by using the shielding factor formulation.
The geometrical diffusion phenomenon presented in \cite{aft} is
explained and clarified, and a universal self-similar asymptotic
behavior is derived.

The organization of the paper is as follows. In Section II, we
will show that all the physics involved in the minimal model can
be rewritten in terms of the electric shielding factor, that
determines the behavior of the charge densities and the local
electric field in the medium. This allows a simple analysis of the
model when written in Lagrangian coordinates. Within this
framework, we perform in Section III, as an illustration of the
Lagrangian formulation, the analysis of planar fronts (without
diffusion). In Section IV, we study the evolution of ionization
fronts in which the initial seed of ionization is such that the
electron density vanishes strictly beyond a certain point for
cylindrical and spherical symmetries. We obtain precise power laws
for both the velocity of the moving fronts and their amplitude. In
Section V, we analyze the special features that appear if the
initial seed of ionization is not completely localized but the
charge densities slowly decrease along the direction of
propagation. For curved geometries, this initial distribution
gives rise to a new diffusion-type behavior that we call
geometrical diffusion. A universal self-similar asymptotic shape
of the fronts is predicted and observed. In Section VI, we
establish our conclusions.

\section{Electric shielding factor}
In this section we will reformulate the problem of the evolution
of streamer fronts in the minimal model by introducing a new
quantity called the electric shielding factor, as in \cite{aft}.
The equation describing the evolution of the shielding factor
makes easier the study of curved ionization fronts.

We begin with a brief discussion about the consequences of
neglecting the magnetic effects in the minimal streamer model. In
this model, it is assumed that the magnetic field effects are
negligible, in a first approximation, because (i) the fluid
velocity of the electrons is much smaller than the velocity of
light, and (ii) the initial magnetic field is zero. Strictly
speaking, if the magnetic field is zero in the evolution of the
ionization wave, then Faraday's law implies that the electric
field is conservative (i.e. $\nabla \times {\bf E} =0$). This
means that, in cases in which the evolution of the ionization wave
is symmetric (planar, cylindrical or spherical), the electric
field would evolve according to this symmetry, so one can write
\begin{equation}
{\bf E} ({\bf r},t) = {\bf E}_{0} ({\bf r}) u({\bf r},t),
\label{relEu}
\end{equation}
where ${\bf E}_{0} ({\bf r})$ is the initial electric field (that
is conservative since it is created by an applied potential
difference) and $u({\bf r},t )$ is some scalar function with the
same symmetry as the initial electric field, since
\begin{equation}
0 = \nabla \times {\bf E}  = {\bf E}_{0} \times \nabla u ,
\label{poisson3}
\end{equation}
and therefore $\nabla u$ is parallel to ${\bf E}_{0}$.
Consequently, the relation (\ref{relEu}) assures that the magnetic
field will always be zero if the initial magnetic field is zero
and it is a direct consequence of the hypothesis of the minimal
streamer model. However, experimental observations indicate that
streamers can change their direction while evolving, suggesting
that the ionization process can be non-symmetric in some cases and
that the local magnetic field may play a role, pointing to future
modifications of the model. Nevertheless, as a first approach to
the problem, we will consider here symmetric situations. In these
cases, the minimal model can be applied for the streamer
evolution, and the relation (\ref{relEu}) is strictly correct.

The above discussion leads to an unexpected consequence on the
minimal streamer model: the quantity $u$ defined in (\ref{relEu})
determines completely, without any physical approximation, the
electric field and the particle densities during the evolution of
the ionization wave if the diffusion is neglected. If we take the
diffusion coefficient $D$ equal to zero, the minimal streamer
model given by equations (\ref{1})-(\ref{ft}) can be rewritten as
\begin{eqnarray}
\frac{\partial n_{e}}{\partial t} &=& \nabla \cdot \left( n_{e}
{\bf E} \right) +
n_{e} |{\bf E}| e^{-1/|{\bf E}|}, \label{elect} \\
\frac{\partial n_{p}}{\partial t} &=& n_{e} |{\bf E}| e^{-1/|{\bf E}|}, \label{ion} \\
\nabla \cdot {\bf E} &=& n_{p} - n_{e}. \label{gauss}
\end{eqnarray}
Subtracting equation (\ref{elect}) from (\ref{ion}), we obtain
\begin{equation}
\frac{\partial}{\partial t} \left( n_{p}-n_{e} \right) = - \nabla
\cdot \left( n_{e} {\bf E} \right) . \label{paso1}
\end{equation}
By taking the time derivative in equation (\ref{gauss}), we obtain
\begin{equation}
\frac{\partial}{\partial t} \nabla \cdot {\bf E} =
\frac{\partial}{\partial t} \left( n_{p}-n_{e} \right) ,
\label{paso2}
\end{equation}
and hence, using (\ref{paso1}), we get
\begin{equation}
\nabla \cdot \left( \frac{\partial {\bf
E}}{\partial t} + n_{e} {\bf E} \right) =0. \label{paso3}
\end{equation}
Since the electric current is given by $n_{e} {\bf E}$, expression
(\ref{paso3}) is simply the divergence of Amp\`ere's law applied
to our case, with the right hand side being the divergence of the
curl of the magnetic field, which is always zero. In the
particular situation in which the curl of the magnetic field in
the gas is negligible, as it occurs in the framework of the
minimal model for symmetric situations (as discussed above), this
expression can be also written as
\begin{equation}
\frac{\partial {\bf E}}{\partial t} + n_{e} {\bf E} =0. \label{paso4}
\end{equation}
This is a linear first-order ordinary differential equation for
the electric field, so that it can be trivially integrated to give
\begin{equation}
{\bf E} ({\bf r}, t) = {\bf E}_{0} ({\bf r}) \exp{ \left( -
\int_{0}^{t} d\tau n_{e} ({\bf r}, \tau ) \right) }, \label{paso5}
\end{equation}
which supplies the local electric field ${\bf E}$ in terms of the
initial electric field ${\bf E}_{0}$ multiplied by the electron
density $n_{e}$ integrated in time. The physical behavior of the
electric field screened by a charge distribution suggests that an
important new quantity can be defined as
\begin{equation}
u({\bf r},t) =\exp{ \left( - \int_{0}^{t} d\tau n_{e} ({\bf r},
\tau ) \right) }, \label{defu}
\end{equation}
so that the relation (\ref{relEu}) is re-obtained. This means
that, when $u$ is determined in a particular situation, the
electric field ${\bf E}$ is known. Moreover, using equations
(\ref{defu}) and (\ref{gauss}), we obtain that the particle
densities are also determined by $u$ and the initial condition
${\bf E}_{0} ({\bf r})$ for the electric field, through
\begin{eqnarray}
n_{e} ({\bf r}, t) &=& -\frac{1}{u({\bf r},t)} \frac{\partial
u({\bf r},t)}{\partial t}, \label{relsigmau} \\
n_{p}({\bf r}, t) &=& -\frac{1}{u({\bf r},t)} \frac{\partial
u({\bf r},t)}{\partial t} + \nabla \cdot \left( {\bf E}_{0} ({\bf
r}) u({\bf r},t) \right). \label{relrhou}
\end{eqnarray}
Equation (\ref{relEu}) reveals clearly the physical role played by
the function $u({\bf r},t)$ as a factor modulating the electric
field ${\bf E} ({\bf r}, t)$ at any time. For this reason, $u$ can
be termed shielding factor and determines a screening length that
depends on time. This is a kind of Debye's length which moves with
the front and leaves neutral plasma behind it \cite{aft}. As the
shielding factor determines the particle densities, equation
(\ref{poisson3}) implies that the particle densities have the same
symmetry as the initial electric field.

The definition of the shielding factor $u$ and the mathematical
treatment explained above reduce the problem of evolution of
charged particle densities and electric field in the gas to a
simpler one: to find equations and conditions for the shielding
factor $u({\bf r},t)$ from equations and conditions for the
quantities ${\bf E}$, $n_{e}$ and $n_{p}$. Substituting equations
(\ref{relEu})--(\ref{relrhou}) into the original model equation
(\ref{elect})--(\ref{gauss}), we find
\begin{equation} \frac{\partial}{\partial t} \left(
\frac{1}{u} \frac{\partial u}{\partial t} - \nabla \cdot \left( {\bf E}_{0} u \right) \right) = |{\bf E}_{0}|
\frac{\partial u}{\partial t} e^{-1/(|{\bf E}_{0}| u)}, \label{evolu1}
\end{equation}
where $|{\bf E}_{0}|$ is the modulus of the initial electric field
${\bf E}_{0}$. The last term in this expression can be written as
\begin{equation}
|{\bf E}_{0}| \frac{\partial u}{\partial t} e^{\frac{-1}{|{\bf
E}_{0}| u}} = \frac{\partial}{\partial t} \int_{0}^{|{\bf
E}_{0}|u} e^{-1/s} ds, \label{evolu2}
\end{equation}
so that
\begin{equation}
\frac{\partial}{\partial t} \left( \frac{1}{u} \frac{\partial u}{\partial t} - \nabla \cdot \left( {\bf E}_{0} u
\right) \right) = \frac{\partial}{\partial t} \int_{0}^{|{\bf E}_{0}|u} e^{-1/s} ds. \label{evolu3}
\end{equation}
This equation can be integrated once in time to give
\begin{equation}
\frac{1}{u} \frac{\partial u}{\partial t} - \nabla \cdot \left( {\bf E}_{0} u \right) = \int_{0}^{|{\bf E}_{0}|u}
e^{-1/s} ds + G({\bf r}), \label{evolu4}
\end{equation}
where the function $G({\bf r})$ is given by
\begin{equation}
G({\bf r})= \left. \left( \frac{1}{u} \frac{\partial u}{\partial t} - \nabla \cdot \left( {\bf E}_{0} u \right) -
\int_{0}^{|{\bf E}_{0}|u} e^{-1/s} ds \right) \right|_{t=0} . \label{evolu5}
\end{equation}
The initial conditions for $u$ and $\partial u/\partial t$ can be
easily related to initial conditions for particle densities using
(\ref{relEu}) and (\ref{relsigmau}). The results are
\begin{eqnarray}
\left. u \right|_{t=0} &=& 1, \label{iniu} \\
\left. \frac{\partial u}{\partial t} \right|_{t=0} &=& - n_{e0}
({\bf r}) , \label{iniut}
\end{eqnarray}
where $n_{e0} ({\bf r})$ is the initial value of $n_{e} ({\bf
r},t)$. Then,
\begin{equation}
G({\bf r})= - n_{e0} ({\bf r}) - \nabla \cdot {\bf E}_{0} - \int_{0}^{|{\bf E}_{0}|} e^{-1/s} ds , \label{evolu6}
\end{equation}
which, if $n_{p0} ({\bf r})$ is the initial value of the
dimensionless ion density, can also be written as
\begin{equation}
G({\bf r})= - n_{p0} ({\bf r}) - \int_{0}^{|{\bf E}_{0}|} e^{-1/s} ds . \label{evolu7}
\end{equation}
As a consequence, the evolution of $u({\bf r},t)$ is given by
\begin{eqnarray}
\frac{1}{u} \frac{\partial u}{\partial t} &=& \nabla \cdot \left( {\bf E}_{0} u \right)
 - n_{p0} ({\bf r}) -
\int_{|{\bf E}_{0}|u}^{|{\bf E}_{0}|} e^{-1/s} ds ,
\label{evolufinal1} \\
u ({\bf r},0) &=& u_{0} ({\bf r}) = 1, \label{evolufinal2}
\end{eqnarray}
with appropriate boundary conditions which should be imposed
depending on the particular physical situations one wishes to
consider. Note that we have written the complete minimal model
(with $D=0$) in one single equation for the shielding factor $u$.
All the physics in the minimal model is contained in the evolution
equation (\ref{evolufinal1}). The shielding factor is related to
charged particle densities and electric field through expressions
(\ref{relEu}), (\ref{relsigmau}) and (\ref{relrhou}). This
formulation (\ref{evolufinal1}) allows us a much simpler analysis
than the original one, and will provide us with some insight into
some unsolved problems on streamer formation.

\section{The planar case and the Lagrangian formulation}

In this section, we will use the shielding factor in order to find
the main features of planar anode-directed ionization fronts
without diffusion. This is a very simple way of testing the
usefulness of the new formulation for further generalization, as
the results can be compared with that found in \cite{Ute} using a
different approach.

\subsection{Lagrangian formulation}

We consider an initial experimental situation as follows. Two
infinite planar plates are situated at $z=0$ and $z=d$
respectively ($z$ is the vertical axis). The space between the
plates is filled with a non-attaching gas like Nitrogen. A
stationary electric potential difference $V_{0}$ is applied to
these plates, so that $V(d)-V(0)=V_{0} > 0$. To initiate the
avalanche, an initial neutral seed of ionization is set at the
cathode, so that $n_{e0} (z) =n_{p0} (z) = \rho_{0} (z)$. We study
the evolution of negative ionization fronts towards the anode at
$z=d$.

\begin{figure}
\centering
\includegraphics[width=0.45\textwidth,height=0.35\textwidth]{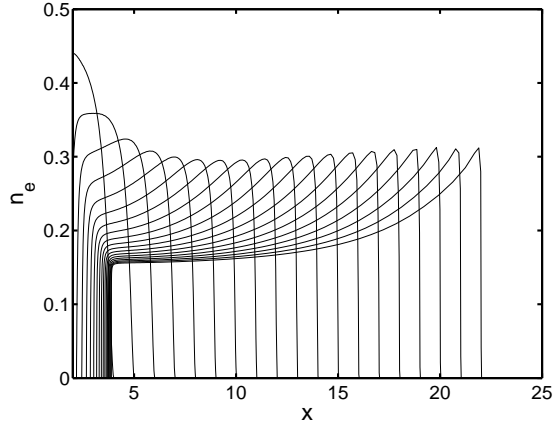}
\caption{Electron density $n_{e}$ of a planar ionization wave for
fixed time intervals vs coordinate $x=z/E_{0}$, in which $z$ is
the propagation direction (from the negative to the positive
plate) and $E_{0}$ is the modulus of the initial electric field
(that is constant) between the plates. The initial data is a
compactly supported neutral seed of ionization near the negative
plate $x=0$. By using the formulation in terms of the shielding
factor we can see that, after evolution, the ionization wave
converges to a travelling wave of constant velocity and constant
amplitude, called a shock front.} \label{fig1}
\end{figure}

As the applied potential is constant, the initial electric field
${\bf E}_{0}$ between the plates results in
\begin{equation}
{\bf E}_{0} = - E_{0} {\bf u}_{z}, \; E_{0} = \frac{V_{0}}{d}.
\label{plane2}
\end{equation}
It is useful for the computations to define the coordinate $x$ as
\begin{equation}
x = \frac{z}{E_{0}}, \label{plane2a}
\end{equation}
so that the evolution of the shielding factor $u$
(\ref{evolufinal1}) is given as the solution of the equations
\begin{eqnarray}
\frac{\partial u}{\partial t} + u \frac{\partial u}{\partial x} &=& - u \rho_{0} (x) - u \int_{E_{0}u}^{E_{0}}
e^{-1/s} ds ,
\label{plane3} \\
u (x,0) &=& u_{0} (x) = 1. \label{plane4}
\end{eqnarray}
This is a typical Burgers type equation with an integral term.
Hence, we can use some of the classical techniques developed to
deal with this equation. In particular, we can integrate along
characteristics and transform equation (\ref{plane3}) into the
system
\begin{eqnarray}
\frac{dx}{dt} &=&u ,  \label{plane8a} \\
\frac{du}{dt} &=&- \rho_{0}(x)u - u \int_{E_{0}u}^{E_{0}} e^{-1/s}
ds,  \label{plane8b}
\end{eqnarray}
which yields a Lagrangian formulation of the problem. The
solutions of this dynamical system with initial data given by
$x(0)=x_{0}$, $u(0)=1$ allows us to compute the profiles for
$u(x,t)$ at any time. Then using equations (\ref{relEu}),
(\ref{relsigmau}) and (\ref{relrhou}), it is possible to trace the
profiles of the electric field or the charge densities at
different times. This has been done in Fig.~\ref{fig1}, in which
the electron density $n_{e}$ is plotted as a function of the
coordinate $x=z/E_{0}$. We have chosen a neutral initial seed of
ionization sufficiently localized near the negative plate, i.e.
the electron and positive ion densities are initially equal and,
moreover, they vanish beyond a certain point in the $x$ axis
(mathematically, this situation is described by saying that the
initial condition is of compact support). After evolution, the
electron density converges to a travelling wave, as can be seen in
Fig.~\ref{fig1}. This travelling wave has a constant propagation
velocity and a constant amplitude, as can be seen in the figure,
and it is a shock front. This shock front appears only if the
initial condition is of compact support, as we will see later.

\subsection{Analytical computations}

The fact that, in the planar case, the integral term in equation
(\ref{plane3}) does not depend explicitly on $x$ has an
interesting consequence: the velocity of the front can be computed
directly from the equation for the shielding factor and it is
completely determined by the initial condition. Let us assume that
the initial density $\rho_{0} (x)$ (for both electron and positive
ion densities) decays sufficiently fast as $x$ goes to infinity.
Then we can neglect the term $u \rho_{0}(x)$ in (\ref{plane3}) for
$x \gg 1$ and look for a solution of the resulting equation in the
form
\begin{equation}
u(x,t)= f(\xi = x - c t). \label{plane9}
\end{equation}
Using the equation (\ref{plane3}) for the evolution of the
shielding factor with this approximations, we obtain the
differential equation
\begin{equation}
\frac{df}{d\xi} = \frac{f v(f)}{c -f} ,\label{plane10}
\end{equation}
where the quantity $v(f)$ is given by
\begin{equation}
v (f) = \int_{E_{0}f}^{E_{0}} e^{-1/s} ds. \label{plane11}
\end{equation}
The front of the wave is localized, at a given time, in points in
which $f \simeq 1$. In these points, up to first order in $f$, the
quantity $v(f)$ results in $v(f) \simeq E_{0} e^{-1/E_{0}} (1-f)$.
Inserting these approximations into equation (\ref{plane10}), we
get
\begin{equation}
\frac{df}{d\xi} = \frac{E_{0} e^{-1/E_{0}}}{c-1} (1-f).
\label{plane13}
\end{equation}
By integrating this expression and using (\ref{relsigmau}) to
obtain the electron density, it can be easily seen that physically
acceptable solutions of this equation (i.e. positive value of the
electron density in all points), correspond only to values of $f$
given by $f \leq 1$ for large $\xi$. From (\ref{plane13}), these
physical solutions appear only if $c \geq 1$. So that the velocity
$c_{z}$ of propagation of the front, in the original $z$
coordinate, satisfies
\begin{equation}
c_{z} \geq E_{0}, \label{plane14}
\end{equation}
in agreement with the result found in \cite{Ute}.

\begin{figure}
\centering
\includegraphics[width=0.45\textwidth,height=0.35\textwidth]{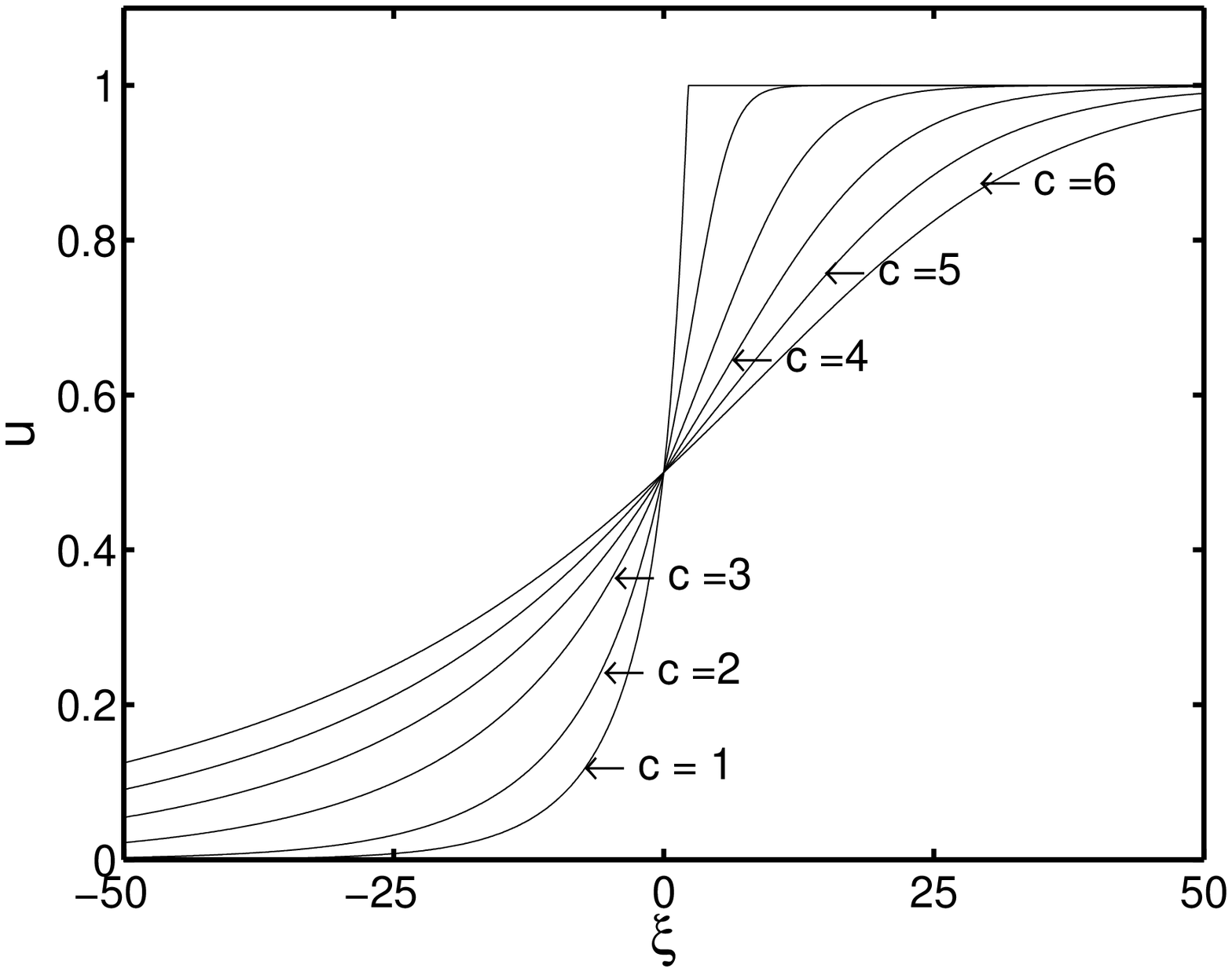}
\includegraphics[width=0.45\textwidth,height=0.36\textwidth]{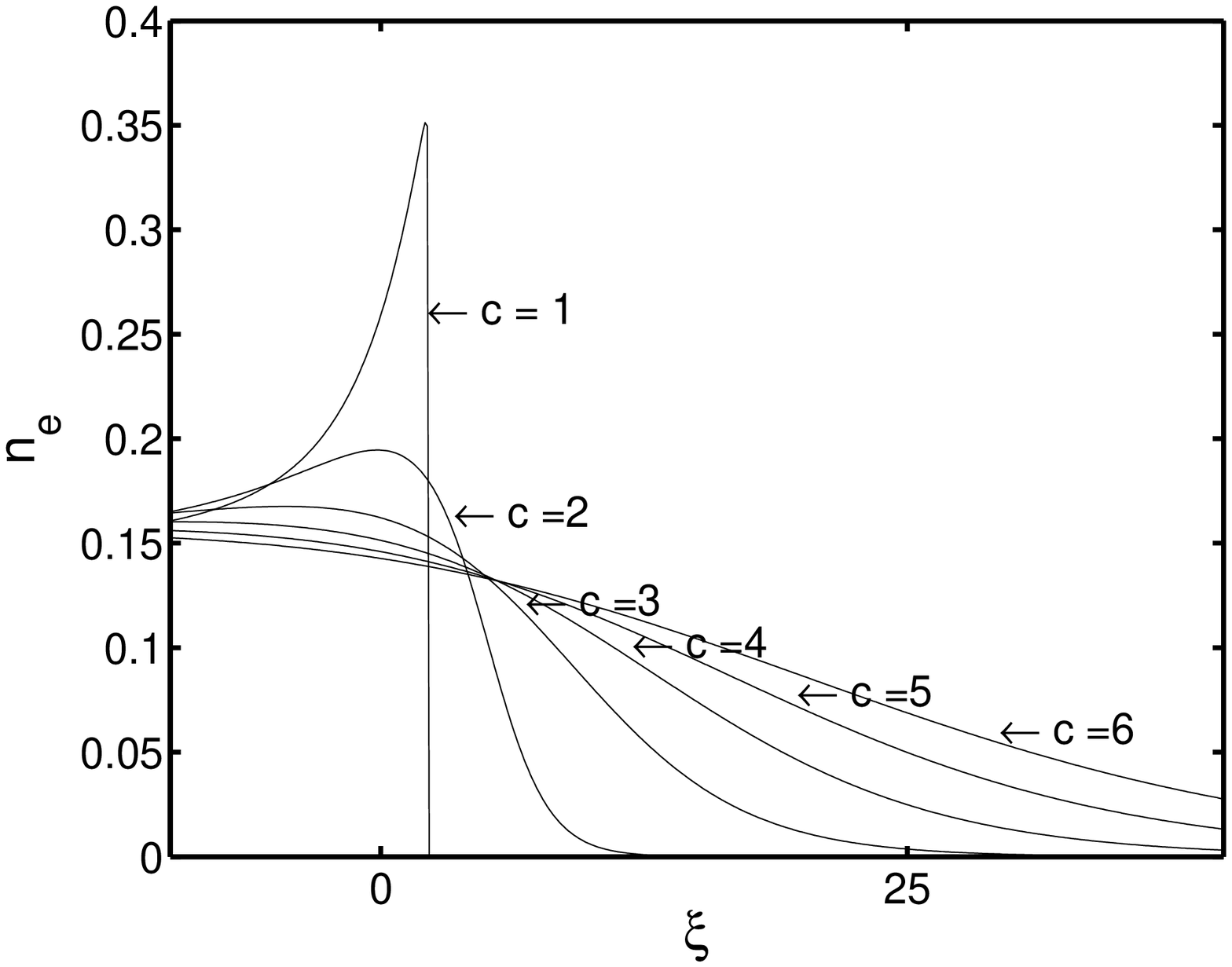}
\caption{(Left) Electric shielding factor $u$, and (Right) electron
density $n_{e}$ of several travelling wave profiles. In the
figures, $\xi = x-ct$ and $c$ takes values from $c=1$ (it
corresponds to a shock front) to $c=6$. The propagation velocity
depends on the way in which the initial electron density behaves
as $x$ goes to $\infty$.} \label{fig2}
\end{figure}

Moreover, by using this formulation it is also possible to link
the asymptotic behavior of the initial condition $n_{e0}$ with the
propagation velocity, so that it will be shown that the initial
condition determines the velocity of the front. Suppose that the
initial condition for the electron density behaves like
$n_{e0}\approx Ae^{-\lambda x}$ as $x \rightarrow \infty$. Then,
the asymptotic behavior of the travelling wave satisfies
\begin{equation}
n_{e}\simeq Ae^{-\lambda \xi } \; \; \mbox{, as } \xi \rightarrow
\infty. \label{plane16}
\end{equation}
Using the relation (\ref{relsigmau}), this means that the
shielding factor $u=f(\xi)$ behaves like
\begin{equation}
f \simeq 1-\frac{A}{\lambda c}e^{-\lambda \xi }\; \; \mbox{, as
}\xi \rightarrow \infty. \label{plane17}
\end{equation}
When this expression is introduced into equation (\ref{plane13}),
the relation
\begin{equation}
\lambda =\frac{E_{0}\exp (-1/E_{0})}{c-1}, \label{plane18}
\end{equation}
appears. This is the way in which the asymptotic behavior of the
initial condition determines the propagation velocity. By using
this link, in Fig.~\ref{fig2} we have plotted several travelling
wave profiles for different values of $c$. In Fig.~\ref{fig2}(left),
the shielding factor has been plotted, and in Fig.~\ref{fig2}(right),
the corresponding electron density, both as a function of $\xi =
x-ct$.

As mentioned above, the minimum value of the propagation velocity
$c$ of the travelling wave is $c=1$ (it corresponds to $c_z
=E_{0}$). For this velocity, a shock front can be clearly seen in
Fig.~\ref{fig2}. Using equation (\ref{plane18}), this shock front
appears when $\lambda \rightarrow \infty$, so that the initial
condition is of compact support, i.e. the initial distribution of
charge vanishes strictly beyond a certain point. In
Fig.~\ref{fig1}, an initial condition fulfilling these
requirements has been chosen and a shock front has appeared as
predicted by (\ref{plane18}). The amplitude of the shock front can
be also obtained from equation (\ref{plane13}) and is given by
\begin{equation}
n_{e} \simeq E_{0} e^{-1/E_{0}}. \label{plane15}
\end{equation}
The rest of profiles in Fig.~\ref{fig2} correspond to travelling
waves in which the velocity $c$ is larger than 1, so the initial
distribution does not have compact support.

\subsection{Accelerated fronts}

The treatment given above could suggest that ionization fronts
always move with constant velocity if the physical setting has
planar symmetry. However, it can be shown that fronts with
constant velocity appear only if the initial conditions for the
particle densities decrease exponentially with the distance from
the cathode. In this context, the compactly supported case is
treated as an exponential decay with infinite argument, as was
done in the previous subsection.

The shielding factor formulation allows us to deduce the existence
of accelerated fronts. This situation occurs when taking an
initial ionization decaying at infinity slower than an
exponential. For instance, by taking
\begin{equation}
n_{e0} \simeq \frac{A}{x^{\alpha }} \;\; \mbox{, \ as } x
\rightarrow \infty \ , \label{accelerated1}
\end{equation}
with $\alpha$ positive. Near the front, we can take $u$ close to
$1$, so that from equation (\ref{plane3}) we get
\begin{equation}
\frac{\partial u}{\partial t} + u \frac{\partial u}{\partial x} =
- \frac{A}{x^{\alpha }}u . \label{accelerated2}
\end{equation}
If we introduce an expression of the form
\begin{equation}
u = F \left( \xi = \frac{x}{t^{1/\alpha}} \right),
\label{accelerated3}
\end{equation}
into equation (\ref{accelerated2}), we obtain
\begin{equation}
- \frac{\xi}{\alpha } \frac{\partial F}{\partial \xi} + t^{(\alpha
- 1)/ \alpha} F \frac{\partial F}{\partial \xi} =
-\frac{A}{\xi^{\alpha}} F , \label{accelerated4}
\end{equation}
which can be approximated by
\begin{equation}
\frac{\xi}{\alpha } \frac{\partial F}{\partial \xi} =
\frac{A}{\xi^{\alpha }}F , \label{accelerated5}
\end{equation}
when $\alpha < 1$ and for large $t$. Hence there might exist
fronts whose position is located at the points
\begin{equation}
\frac{x}{t^{1 / \alpha}} = C , \label{accelerated6}
\end{equation}
implying a superlinear propagation, i.e. an acceleration. This
result illustrates the usefulness of our method, showing an
unexpected behavior of the well studied planar fronts.

\section{Curved symmetries}

When the initial particle distributions or the initial electric
field do not have planar symmetry, the ionization wave behaves
quite differently from what we have described in the previous
section. In particular, the amplitude and the velocity of the
travelling wave are always not constant. The shielding factor
formulation can be applied to those more general curved cases with
few changes with respect to the planar case. We will use this
formulation to treat the cases of cylindrical and spherical
symmetries. In the cylindrical case, we will see that the velocity
of the fronts varies in time as $t^{-1/2}$ and the amplitude of
the front goes as $1/t$ when the initial conditions for the
charged particle densities decay sufficiently fast with the
distance to the cathode. In the spherical case, the velocity goes
as $t^{-2/3}$ and the amplitude varies as $1/t$ as in the
cylindrical case. Both cases can be dealt in a very similar way.

\subsection{Cylindrical symmetry}

First we analyze the case with cylindrical symmetry. We consider
the experimental situation of two cylindrical plates with radius
$r_{0}$ and $r_{1} \gg r_{0}$, respectively. The space between the
plates is filled, as in the planar case, with a non-attaching gas.
A constant potential difference $V_{0}$ is applied to the plates,
so that $V(r_{1})-V(r_{0})=V_{0}> 0$. Then the initial electric
field ${\bf E}_{0}(r)$ between the plates is
\begin{equation}
{\bf E}_{0} (r) = -\frac{B}{r} {\bf u}_{r}, \; \; \; B =
\frac{V_{0}}{\log{(r_{1}/r_{0})}}, \label{cylin2}
\end{equation}
where $B$ is a positive constant and $r$ is the radial coordinate,
ranging from $r_{0}$ to $r_{1}$ . An initial neutral seed of
ionization $\rho_{0} (r)$ with cylindrical symmetry is taken, so
that $n_{e0} (r) =n_{p0} (r) = \rho_{0} (r)$. It is useful to
change the spatial variable $r$ to
\begin{equation}
x=\frac{r^2}{2B}, \label{cylin8}
\end{equation}
so that equation (\ref{evolufinal1}) for the shielding factor $u$
takes the form of the Burgers equation
\begin{equation}
\frac{\partial u}{\partial t} + u \frac{\partial u}{\partial x} = - u \rho_{0} (x) - u
\int_{\sqrt{B/(2x)}u}^{\sqrt{B/(2x)}} e^{-1/s} d s . \label{cylin9}
\end{equation}
Since the integral term in equation (\ref{cylin9}) depends
explicitly on $x$, it is quite convenient to define the variable
$v(x,t)$ through
\begin{equation}
v (x,t) = \int_{\sqrt{B/(2x)}u}^{\sqrt{B/(2x)}} e^{-1/s} d s . \label{cylin10}
\end{equation}
Now, as in the case of planar symmetry, we can
integrate (\ref{cylin9}) along characteristics, transforming this
equation into the system of ordinary differential equations
\begin{eqnarray}
\frac{dx}{dt} &=&u ,  \label{cylin11} \\
\frac{du}{dt} &=&-uv-\rho_{0}(x)u ,  \label{cylin12}
\end{eqnarray}
Given the definition of $v$ in (\ref{cylin10}), by taking its time
derivative, and using (\ref{cylin11}) and (\ref{cylin12}), we
close the above system with the equation
\begin{eqnarray}
\frac{dv}{dt} &=& \frac{\sqrt{B/2}}{2} e^{-\sqrt{2x/B}/u} x^{-3/2}
u^2 \nonumber
\\ &-& \frac{\sqrt{B/2}}{2} e^{-\sqrt{2x/B}} x^{-3/2} u \label{cylin13} \\
&+& \sqrt{B/2} e^{-\sqrt{2x/B}/u} x^{-1/2} \left[ uv+ \rho_{0}(x)u
\right]. \nonumber
\end{eqnarray}
Equations (\ref{cylin11}), (\ref{cylin12}), (\ref{cylin13}),
constitute a Lagrangian description of the problem. This dynamical
system can be solved with appropriate initial conditions
$x(0)=x_{0}$, $u(0)=1$, $v(0)=0$, for any $x_{0}$, allowing us to
obtain the profiles for the function $u(x,t)$ at any time $t$.

\begin{figure}
\centering
\includegraphics[width=0.45\textwidth,height=0.35\textwidth]{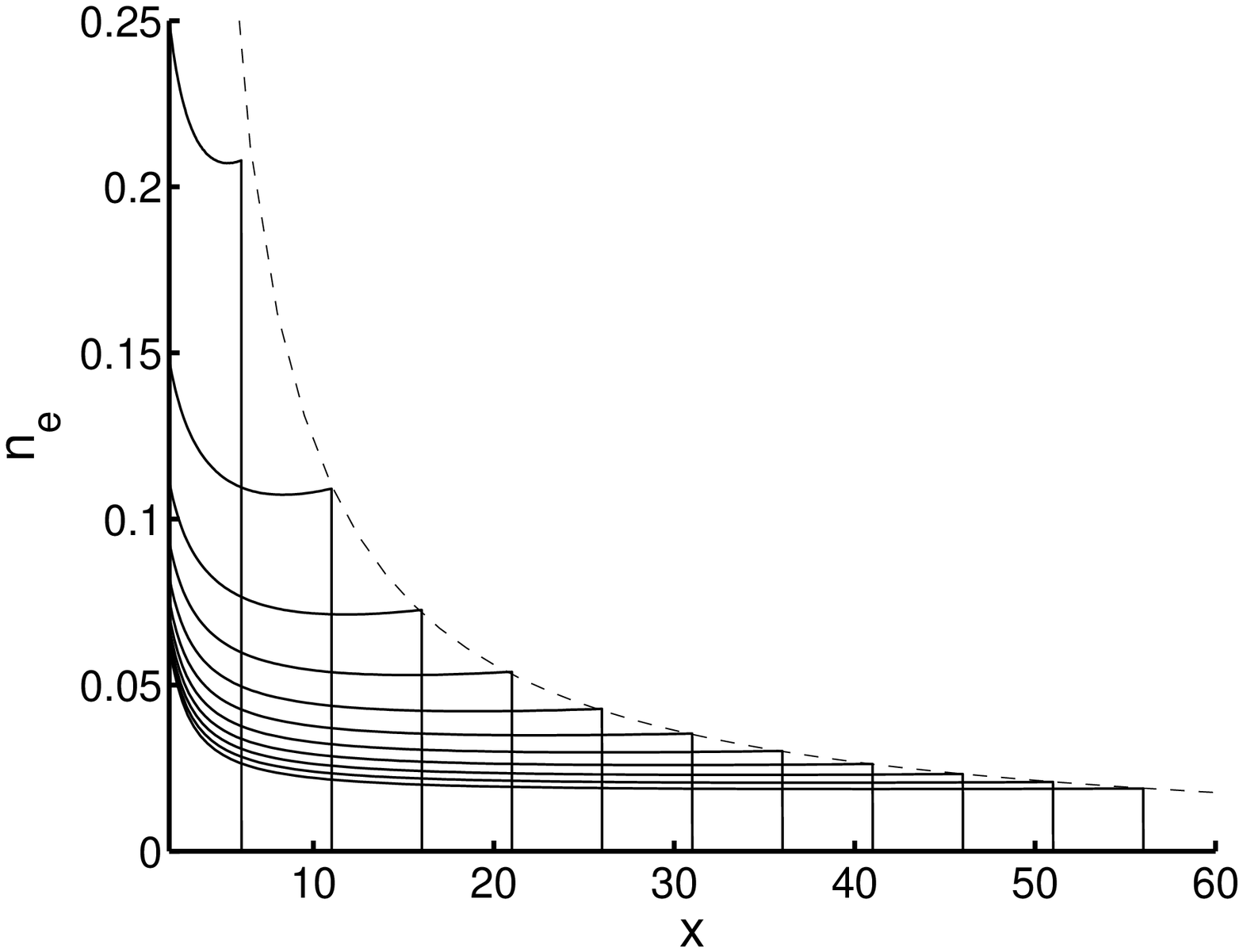}
\includegraphics[width=0.45\textwidth,height=0.35\textwidth]{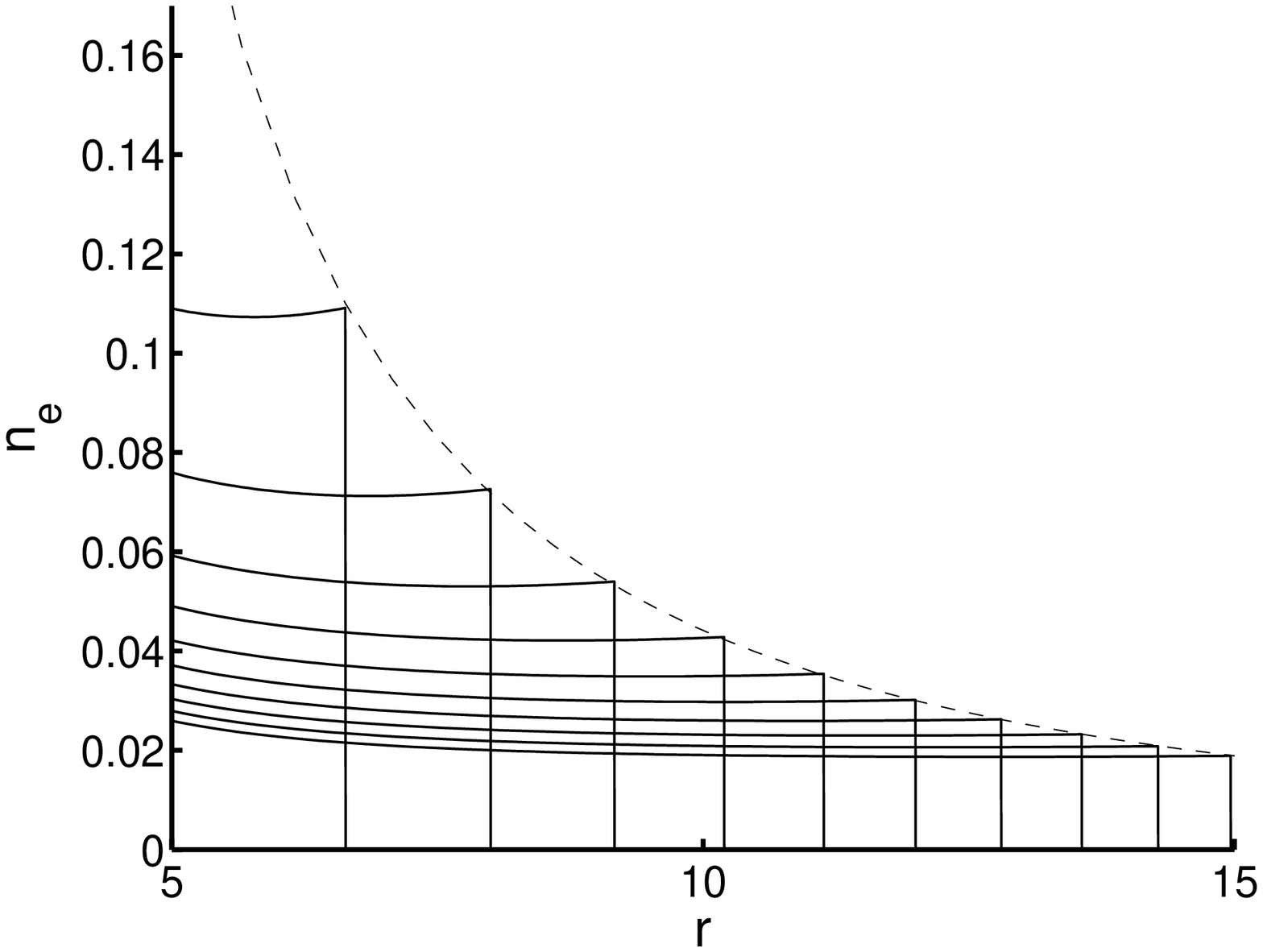}
\caption{(Left) Electron density $n_{e}$ of an ionization wave with
cylindrical symmetry for fixed time intervals vs coordinate
$x=r^{2}/(2B)$, where $r$ is the radial coordinate and $B$ is a
constant related to the initial electric field between the plates.
A compactly supported neutral seed of ionization near the cathode
($x=0$) has been considered as initial condition for the electron
density. The solution converges into a shock front with decaying
amplitude and constant velocity (in terms of the non-physical $x$
coordinate). The dashed line is the analytical prediction for the
amplitude of the shock front. (Right) Same conditions as in (left) but
the electron density has been plotted vs the physical radial
coordinate $r$. The velocity is clearly non-constant, as explained
in the text. The dashed line is the analytical prediction for the
amplitude.} \label{fig3}
\end{figure}

\begin{figure}
\centering
\includegraphics[width=0.45\textwidth,height=0.35\textwidth]{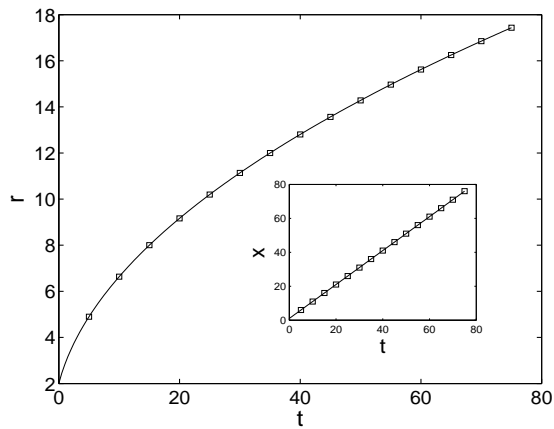}
\caption{Points are the positions $r$ of the shock front shown in
Fig.~\ref{fig3} vs time (in the inset, we plot the positions in
terms of $x=r^{2}/(2B)$). Solid lines are, in both cases, the
analytical predictions for the position of the shock front as a
function of time.} \label{fig3c}
\end{figure}

The solutions of the dynamical system given by equations
(\ref{cylin11}), (\ref{cylin12}), (\ref{cylin13}), depend on the
particular choice of the initial condition for the electron and
positive ion densities. In these equations we have used a neutral
initial condition given by $n_{e0} (r) =n_{p0} (r) = \rho_{0}
(r)$. Consider now the especial case in which the initial
condition $\rho_{0} (r)$ for both densities is compactly
supported, strictly vanishing beyond a certain point. One example
of this behavior is given by a homogeneous thin layer of width
$\delta \ll (r_{1}-r_{0})$ from $r=r_{0}$ to $r=r_{0}+\delta$,
i.e.
\begin{equation}
\rho_{0} (r) = \left\{\begin{array}{ll} \rho_{0} &, \; \; r_{0}<r< r_{0}+\delta \\
0 &, \;\; r_{0}+\delta < r < r_{1} \end{array} \right.
\label{cylin14}
\end{equation}
By using this initial condition, we have plotted in
Fig.~\ref{fig3} the electron density distribution $n_e$
corresponding to a given choice of the physical parameters
$V_{0}$, $\rho_{0}$, $\delta$, $r_{0}$ and $r_{1}$. The electron
density has been calculated from the shielding factor $u$ using
the relation (\ref{relsigmau}) and plotted as a function of $x$ at
fixed time intervals. In Fig.~\ref{fig3}(left), the electron density
has been plotted as a function of coordinate $x$. In
Fig.~\ref{fig3}(right), it has been plotted as a function of the
radial coordinate $r$ with the help of the relation
(\ref{cylin8}). What we can see from the figure is a shock with
decaying amplitude, separating the region with charge and the
region without charge. The numerical data allows us to measure the
velocity of propagation of such front. In Fig.~\ref{fig3c}, we
have plotted the position of the shock $r_f$ as a function of time
$t$. The velocity of propagation is clearly not constant. However,
when we plot the position of the front in terms of $x$, one can
observe the following linear relation (see inset
Fig.~\ref{fig3c}),
\begin{equation}
x_f(t)=t+x_{0}, \label{cylin15}
\end{equation}
which implies, in the original cylindrical variable $r$, an
asymptotic behavior such that the position of the front depends on
time as
\begin{equation}
r_f(t) \simeq \sqrt{2Bt}, \label{cylin16}
\end{equation}
so that the velocity of the front behaves as
\begin{equation}
c_{r} (t) \simeq \sqrt{\frac{B}{2t}}. \label{cylin16b}
\end{equation}
Using (\ref{cylin2}) and (\ref{cylin16}), this result can also be
written as
\begin{equation}
c_{r} \simeq E_{0} (r), \label{cylin16c}
\end{equation}
showing a close similarity with the case (\ref{plane14}) of planar
symmetry. Physically, it means that the shock front moves with the
drift velocity of electrons as it should be expected.

This behavior can also be obtained analytically, as well as the
amplitude of the shock front. This is a considerable advantage of
using the formulation in terms of the shielding factor. To do such
computation, it is useful to write, locally near the front, the
solution for $u$ as
\begin{equation}
u(x,t)=1-a(t)\varphi(\xi), \;\; \xi= x-x_f(t). \label{cylin17}
\end{equation}
This expression can be substituted into equation (\ref{cylin9}).
The computation is simplified if we note that the integral term in
(\ref{cylin9}) is very small when $x \gg 1$. We get
\begin{equation}
a(t)\varphi^{\prime}(\xi)-a(t)\varphi^{\prime}(\xi)x^{\prime}_f(t)-a^{\prime }(t)\varphi(\xi)+a^{2}(t)\varphi
(\xi)\varphi^{\prime }(\xi) \approx 0. \label{cylin18}
\end{equation}
The only way this equation can be satisfied is by choosing
\begin{eqnarray}
x_f(t)&=&t+x_0\;,\label{cylin19} \\
a(t)&=& \frac{\beta}{(t+t_{0})}, \label{cylin20} \\
\varphi (\xi) &=& {\beta}^{-1}(x-t-x_{0}), \label{cylin21}
\end{eqnarray}
where $\beta$ is an arbitrary constant depending on initial
conditions. Equation (\ref{cylin19}) is an analytical proof of the
numerical law (\ref{cylin15}) obtained for the position of the
front in terms of time. To obtain the amplitude of the shock
front, we use relation (\ref{relsigmau}) to compute the electron
density from the asymptotic solution (\ref{cylin17}). What we get
is
\begin{equation}
n_e(x,t) \simeq \left\{\begin{array}{ll}\frac{1}{t+t_0}\;\frac{1+
(x-t-x_0)/(t+t_0)}{1-
(x-t-x_0)/(t+t_0)} &,\; x\le t+x_0\\
0 \; &,\; x>t+x_0
\end{array}
\right. , \label{cylin20b}
\end{equation}
which implies that the amplitude of the front decays with time as
\begin{equation}
n_e(x_f(t),t)=\frac{1}{t+t_0}. \label{cylin21b}
\end{equation}
The analytical curve (\ref{cylin21b}) has been plotted as a dashed
line in Figs.~\ref{fig3}(left) and \ref{fig3}(right). The agreement with
numerical data is seen to be excellent, especially for large
times.

\subsection{Spherical symmetry}

The physical case in which the initial electric field and the
initial particle densities have spherical symmetry shows close
similarities with the cylindrical symmetry case. The shielding
factor formulation for spherical symmetry has been used in
\cite{aft} to analyze a typical corona discharge. In this example,
we have two spherical plates with radius $r_{0}$ and $r_{1} \gg
r_{0}$, in which a potential difference $V(r_{1})-V(r_{0})=V_{0}>
0$ is applied. Note that, in this case, $r$ is the spherical
radial coordinate. The initial seed of ionization is neutral so
that $n_{e0}(r) = n_{p0} (r) = \rho_{0}(r)$, and the initial
electric field ${\bf E}_{0}(r)$ between the plates is
\begin{equation}
{\bf E}_{0} (r) = -\frac{C}{r^2} {\bf u}_{r}, \; C = V_{0} \frac{r_{0} r_{1}}{r_{1}-r_{0}}. \label{spher2}
\end{equation}
Changing the spatial variable $r$ to
\begin{equation}
x=\frac{r^3}{3C}, \label{spher8}
\end{equation}
the evolution for the screening factor takes the form of the
Burgers' equation
\begin{equation}
\frac{\partial u}{\partial t} + u \frac{\partial u}{\partial x} = - u \rho_{0} (x) - u \int_{u\left(\frac{C}{9
x^2}\right)^{1/3} }^{\left(\frac{C}{9 x^2}\right)^{1/3}} e^{-1/s} d s , \label{spher9}
\end{equation}
where $\rho_{0}(x)$ is the initial distribution of charge. This
equation, as in the case of cylindrical symmetry (\ref{cylin9}),
can be integrated along characteristics. The results are very
similar to that of cylindrical symmetry shown above \cite{aft}.
For the case of sufficiently localized initial conditions, when
the initial electron density strictly vanishes beyond a certain
point, there appears a sharp shock with decaying amplitude,
separating the region with charge and the region without charge.
The velocity of propagation of such front is given by the relation
between the position of the front and time: $x_f(t)=t+x_{0}$. This
implies, in terms of the original variable $r$, an asymptotic
behavior
\begin{equation}
r_f(t) \simeq (3C)^{1/3}\;t^{1/3}, \label{spher15}
\end{equation}
for the position of the front. The velocity of the front is then
\begin{equation}
c_{r} (t) \simeq \frac{1}{3} (3C)^{1/3} \; t^{-2/3},
\label{spher15b}
\end{equation}
or, in terms of the initial electric field (\ref{spher2}),
\begin{equation}
c_{r} \simeq E_{0} (r). \label{spher15c}
\end{equation}
The analytical computation of the amplitude and propagation
velocity of the shock can be done, by taking the shielding factor
near the front as $u(x,t)=1-a(t)\varphi(\xi)$, which gives exactly
the same equation found in the cylindrical case (\ref{cylin18}).
The reason of this is that the integral term is neglected in both
cases. However, note that the relation between the coordinate $x$
and the physical radial coordinate is different in each case. The
position of the front then satisfies $x_f(t)=t+x_{0}$ and the
amplitude of the electron density front also decays with the law
$n_e(x_f(t),t)=1/(t+t_0)$. Details and figures showing these
results can be found in \cite{aft}.

\section{Geometrical diffusion and self-similar behavior}

In previous sections we have found that, in the framework of the
minimal streamer model without diffusion, when an initial seed of
ionization is placed near the cathode, a travelling ionization
wave develops towards the anode. The shape and the velocity of
this wave depends on the asymptotic behavior of the initial
particle (electron and positive ion) density. If the initial
particle density is compactly supported, i.e. it vanishes beyond a
certain point, then the travelling wave is a shock front, whose
velocity is equal to the drift velocity of electrons, i.e. equal
to the modulus $E_{0}$ of the initial electric field. This
behavior is found in the case in which the physical situation has
planar, cylindrical or spherical symmetry. In the last two cases,
the velocity, as the initial electric field, is not uniform. With
respect to the amplitude of the electron density of the shock, in
the planar case, it is constant during the evolution, but it
decays as $1/t$ in the curved case (cylindrical or spherical
symmetry).

When the initial particle density is not compactly supported, but
it does decay exponentially with the distance from the cathode,
the shock front does not appear. In the planar case, we have seen
that the velocity of the front is then constant and larger than
the drift velocity of the electrons, and the amplitude is
constant. However, if the initial particle density decay slower
than an exponential with the distance from the cathode,
accelerated fronts might appear.

Now we are going to investigate the especial features that appear
in a case with cylindrical or spherical symmetry when the initial
particle densities are not compactly supported but they decay
exponentially with the distance from the cathode. We will see that
(i) a shock front does not appear, (ii) but a front with an
asymptotic self-similar behavior, (iii) whose velocity depends on
time in a similar way as the velocity of the shock front seen in
the previous section. As a remarkable fact, we will note the
appearance of a new type of diffusion effect, due to the geometry
of the initial physical situation.

\begin{figure}
\centering
\includegraphics[width=0.45\textwidth,height=0.35\textwidth]{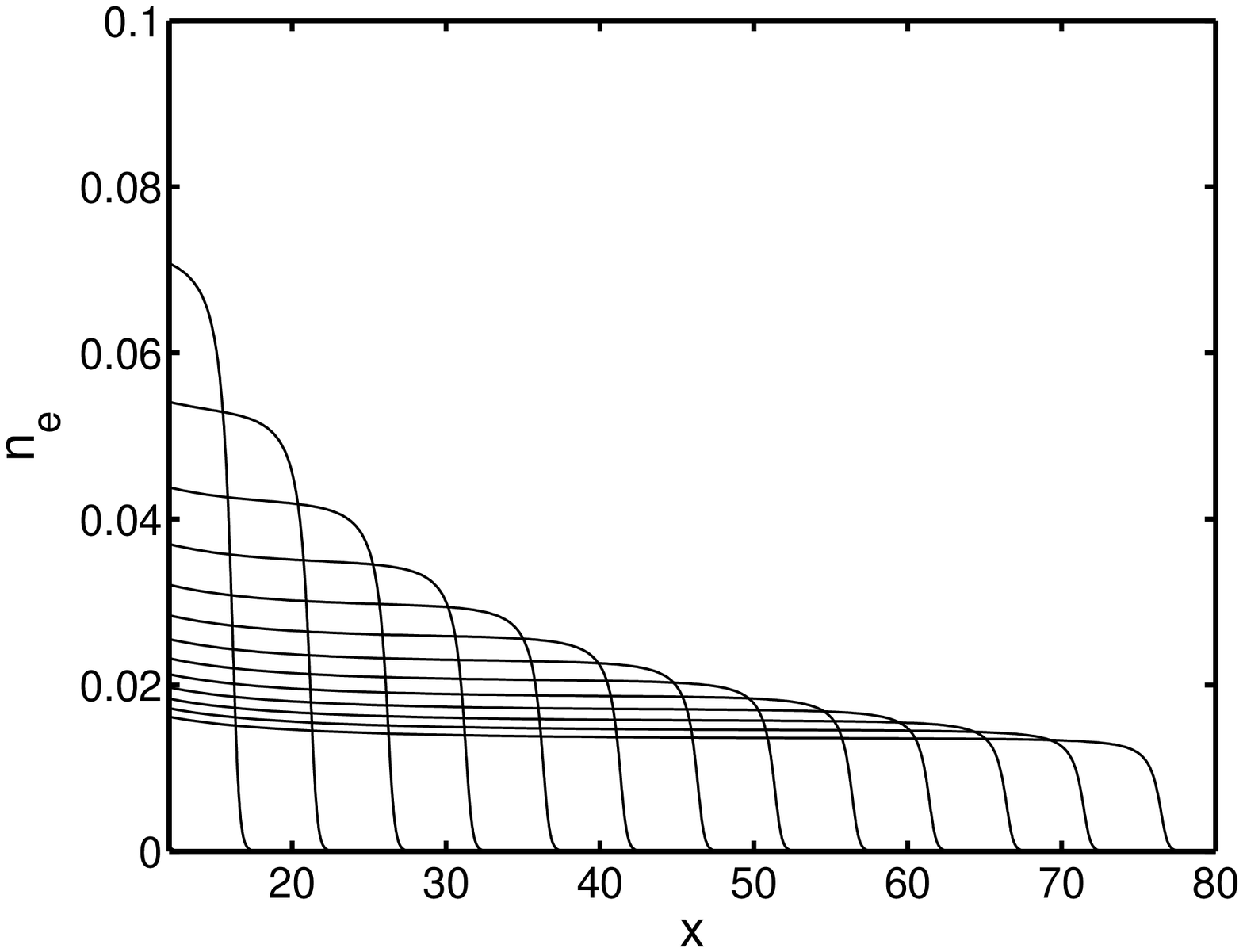}
\includegraphics[width=0.45\textwidth,height=0.35\textwidth]{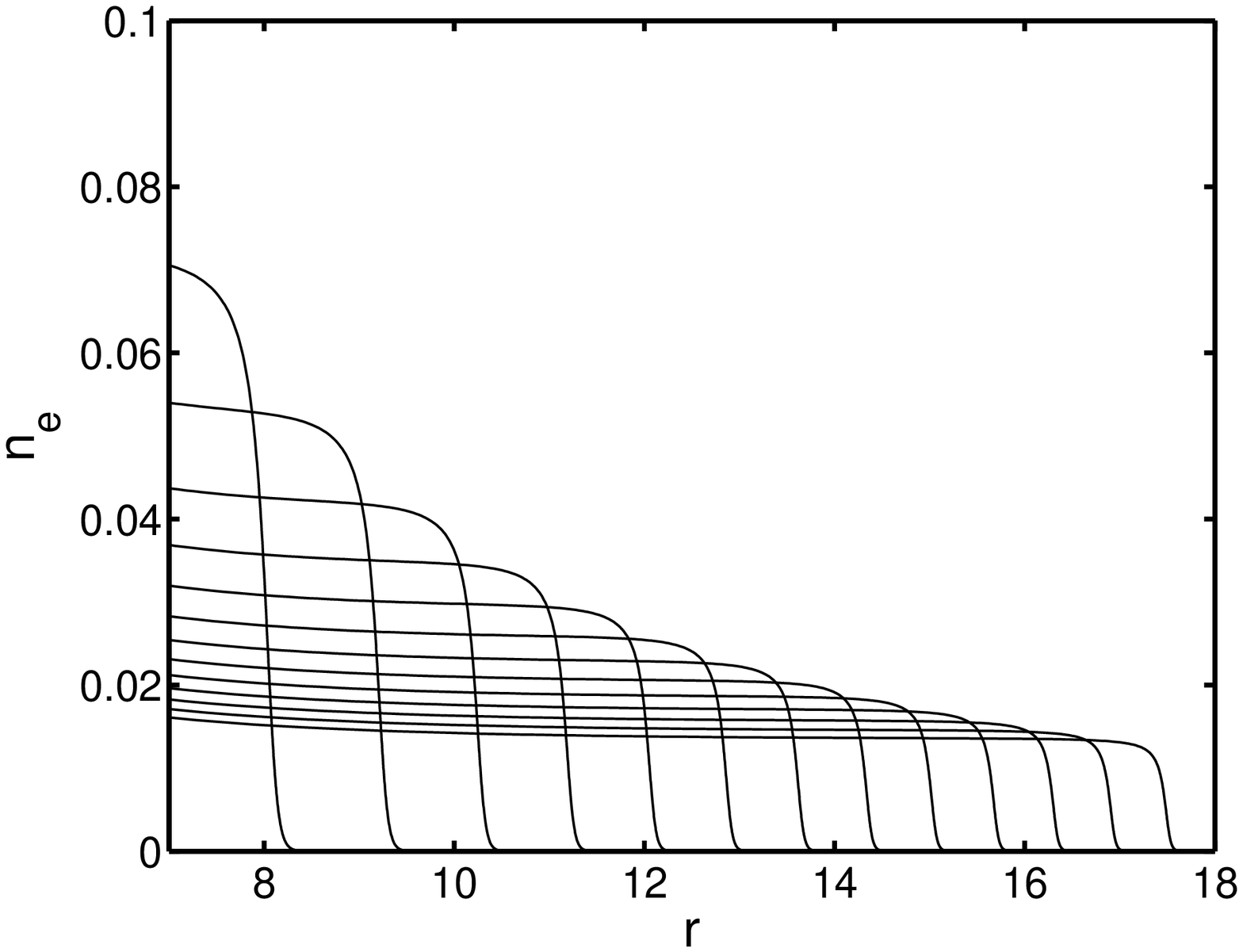}
\caption{(Left) Electron density $n_{e}$ of an ionization wave with
cylindrical symmetry for fixed time intervals vs coordinate
$x=r^{2}/(2B)$, where $r$ is the radial coordinate. An initial
neutral seed of ionization near the cathode ($x=0$) has been
considered such that it exponentially decays with the distance
from the cathode. The solution converges into a spreading front
with constant velocity in terms of $x$. (Right) Same conditions as in
(left) but the electron density has been plotted vs the physical
radial coordinate $r$. The velocity is clearly non-constant in
terms of $r$.} \label{fig4}
\end{figure}

Consider the case of an initial electric field with a cylindrical
symmetry as (\ref{cylin2}). The initial electron and positive ion
densities are equal and have cylindrical symmetry, so that $n_{e0}
(r) =n_{p0} (r) = \rho_{0} (r)$, where $r$ is the radial
coordinate. We use the variable $x=r^{2}/(2B)$ as in the previous
section. Now we take an initial neutral charge distribution that
is not compactly localized, for example given by
\begin{equation}
n_{e0}(x)=n_{p0}(x)=\rho_{0}(x)\sim e^{-\lambda x} , \; \; x \gg
1. \label{cylin22}
\end{equation}
We can solve this problem numerically, integrating along
characteristics the dynamical system given by equations
(\ref{cylin11}), (\ref{cylin12}) and (\ref{cylin13}). The obtained
electron density appears in Fig.~\ref{fig4}, shown in constant
time intervals, proving that the shock front does not appear. In
Fig.~\ref{fig4}(left), we plot the electron density vs $x$ for
different times. What we see is a travelling wave with increasing
thickness and, remarkably, the center of this front moves with
constant velocity in the coordinate $x$ in the same way as the
shock front does. This means that the velocity of the front center
has a similar behavior $c_{r} \sim t^{-1/2}$ that the velocity of
the shock front that we analyzed in the previous section. In
Fig.~\ref{fig4}(right), we plot the electron density vs the radial
coordinate $r$.

\begin{figure}
\centering
\includegraphics[width=0.45\textwidth,height=0.35\textwidth]{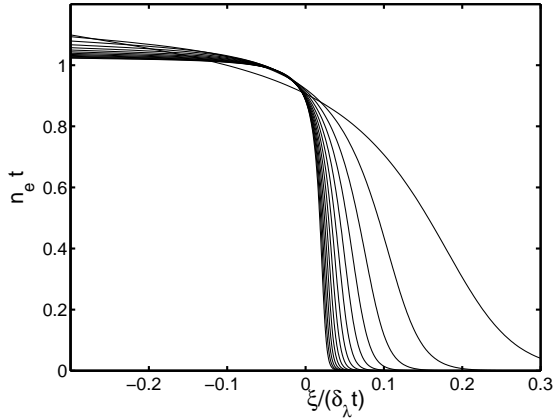}
\caption{Asymptotic behavior of the electron density shown in
Fig.~\ref{fig4}. We plot the quantity $n_{e}t$ vs $\xi /(
\delta_{\lambda}t)$, and a self-similar behavior is apparent. The
front spreads in time, showing a new type of diffusion effect
completely due to the geometry of the initial field distribution.}
\label{fig4c}
\end{figure}

By using the shielding factor formulation, we can prove that the
asymptotic local behavior of the electron density near the front
of the travelling wave is self-similar. In order to show this
property, we introduce
\begin{equation}
u(x,t)=1-g(x,t), \label{gd1}
\end{equation}
in the equation (\ref{cylin9}) describing the evolution of the
shielding factor. As we are analyzing the asymptotic behavior, we
can take $x\gg 1$ near the front. Keeping the main order terms we
obtain the equation
\begin{equation}
\frac{\partial g}{\partial t}+(1-g)\frac{\partial g}{\partial
x}=0, \label{gd2}
\end{equation}
which can be written as
\begin{equation}
\frac{\partial g}{\partial t}-g\frac{\partial g}{\partial \xi}=0,
\label{ecuf}
\end{equation}
if we define $\xi = x-t$. Equation (\ref{ecuf}) is a Burgers
equation whose solution $g$ is such that it is constant along the
curves given by
\begin{equation}
\frac{d\xi }{dt}=g . \label{gd3}
\end{equation}
If $g$ varies slowly in some region of size $\delta_{\lambda}$ in space, 
one can consider
$g=G$ at that region, $G$ being a constant. An intermediate
asymptotic regime is then obtained, such that it is constant along
$\xi =Gt$ and therefore we can assume
\begin{equation}
g(x,t) \simeq g\left( \frac{\xi }{\delta _{\lambda }t}\right) .
\label{gd4}
\end{equation}
Hence the asymptotic behavior of the electron density is given by
\begin{equation}
n_{e} = -\frac{1}{u} \frac{\partial u}{\partial t}
 \simeq \frac{1}{t}\frac{\xi /(\delta _{\lambda }t)}{1+g \left(
 \xi/(\delta _{\lambda }t) \right)} \frac{d}{d \xi} g \left(
 \xi/(\delta _{\lambda }t) \right). \label{gd5}
\end{equation}
Consequently, the asymptotic local behavior of the electron
density near the front is self-similar, given by
\begin{equation}
n_e (x,t) \simeq \frac{1}{t} f\left( \frac{\xi}{\delta_{\lambda
}t}\right), \label{cylin23}
\end{equation}
in which $\xi = x-t$, and $f$ is some universal self-similar
profile. Hence the front presents a typical thickness given by
\begin{equation} \xi_c \simeq \delta _{\lambda }t .
\label{cylin24}
\end{equation}
This result can be seen in Fig.~\ref{fig4c}, in which the
self-similar character of the asymptotic local behavior is shown.
The consequence of this result is clear: even neglecting
diffusion, the front spreads out linearly in time when the initial
condition for the particle densities decreases exponentially with
the distance from the cathode. This is a new and remarkable
feature, first considered in \cite{aft} and explained here, of the
curved geometry. It is a diffusive behavior of the solutions of
the minimal streamer model caused by the geometry of the electric
field. It has been termed {\it geometrical diffusion}.

In the spherical case, the same behavior is found, with the only
difference being that the $x$ coordinate is related to the radial
coordinate $r$ in a different way. We conclude than geometrical
diffusion is a universal behavior.

\section{Conclusions}

The conclusions of this paper are the following. We have made a
thorough study of the properties and structure of anode-directed
ionization fronts without diffusion, based on a minimal streamer
model for non-attaching gases. This model includes impact
ionization processes as source terms. The role played by the
condition that the magnetic effects in the streamer discharges are
neglected has been discussed. As a consequence, it has been proved
that an electric shielding factor can be defined, and the physical
quantities can be expressed as a function of it.

A Burgers type equation is obtained for the evolution of the
electric shielding factor. Thus, the analytical and numerical
study of the ionization fronts can be performed by using a
Lagrangian formulation. The power of this formulation makes it
easier to treat the cases of the non-homogeneous initial electric
field in electric discharges with curved symmetries.

We have applied this new formulation to the discharge between
planar and curved electrodes (with cylindrical and spherical
symmetries). When an initial seed of ionization is placed near the
cathode, a travelling wave develops towards the anode. The shape
and the velocity of this wave depends on the asymptotic behavior
of the initial charged particle densities. If the initial density
is compactly supported, then the travelling wave is a shock front,
whose velocity is equal to the drift velocity of electrons. This
behavior had been predicted for the planar case \cite{Ute}, but we
have found that a similar situation takes place in the case that
the physical situation has cylindrical or spherical symmetries. We
have derived power laws for the velocity and the amplitude of the
shock fronts in the cases of curved symmetry. When we have
cylindrical symmetry, the velocity of the shock front behaves as
$t^{-1/2}$, and the amplitude behaves as $t^{-1}$. In the
spherical case, the velocity of the shock front behaves as
$t^{-2/3}$ and the amplitude goes as $t^{-1}$.

When the initial particle density is not compactly supported, but
decays exponentially with the distance from the cathode, the shock
front does not appear. In the planar case, we have seen that the
velocity of the front is then constant and larger than the drift
velocity of the electrons, as predicted in \cite{Ute}. However, if
the initial particle density decay slower than an exponential with
the distance from the cathode, accelerated fronts appear.

In the cases in which the physical situation has cylindrical or
spherical symmetries and the initial ionization seed decays
exponentially fast to infinity, we have seen that the velocity
follows the same power laws as the compactly supported case.
However, the structure of the travelling wave is rather different.
We have proved that the asymptotic behavior of the charged
particle densities is self-similar. Even if the diffusion has not
been considered, the front spreads out linearly in time. This is a
remarkable feature, first considered in \cite{aft} and explained
here. It is a diffusive behavior of the solutions of the minimal
streamer model caused by the geometry of the electric field, and
we have called it geometrical diffusion.

Our analysis opens the way to consider geometrical effects in the
stability of ionization fronts.

\acknowledgments

This paper has been partially supported by the Spanish Ministry of
Science and Technology grant BFM2002-02042, and by the Universidad
Rey Juan Carlos grant PPR-2004-38.

\end{document}